\documentclass[preprint,12pt,3p]{elsarticle}
\usepackage{algorithm}
\usepackage{algorithmic}
\usepackage{mathtools}
\usepackage{amssymb}
\DeclarePairedDelimiter{\size}{\lvert}{\rvert}

\DeclarePairedDelimiter{\poly}{\textrm{poly}(}{)}
\newtheorem{proposition}{Proposition}
\newtheorem{lemma}{Lemma}
\newtheorem{theorem}{Theorem}

\newdefinition{definition}{Definition}
\newproof{pot}{Proof of Theorem~\ref{theo-main}}

\journal{Discrete Mathematics}

\begin{document}

\begin{frontmatter}

\title{An Indexing for Quadratic Residues Modulo $N$ and a Non-uniform Efficient Decoding Algorithm}

\author[label1]{Nicollas M. Sdroievski\corref{cor1}}

\address[label1]{Department of Computer Science, Federal University of Paran\'a,  81531-980, Curitiba, Brazil}
\cortext[cor1]{I am corresponding author}

\ead{nmsdroievski@inf.ufpr.br}

\author[label1]{Murilo V. G. da Silva}
\ead{murilo@inf.ufpr.br}

\author[label1]{Andr\'e L. Vignatti}
\ead{vignatti@inf.ufpr.br}

%TODO mudou
\begin{abstract}
An \emph{indexing} of a finite set $S$ is a bijection $D : \{1,...,|S|\} \rightarrow S$. We present an indexing for the set of quadratic residues modulo $N$ that is decodable in polynomial time on the size of $N$, given the factorization of $N$. One consequence of this result is a procedure for sampling quadratic residues modulo $N$, when the factorization of $N$ is known, that runs in strict polynomial time and requires the theoretical minimum amount of random bits (i.e., $\log{(\phi(N)/2^r)}$ bits, where $\phi(N)$ is Euler's totient function and $r$ is the number of distinct prime factors of $N$). A previously known procedure for this same problem runs in expected (not strict) polynomial time and requires more random bits.
\end{abstract}

\begin{keyword}
%% keywords here, in the form: keyword \sep keyword
coding theory \sep theory of computation \sep computational complexity \sep quadratic residues \sep indexing
%% MSC codes here, in the form: \MSC code \sep code
%% or \MSC[2008] code \sep code (2000 is the default)
\end{keyword}

\end{frontmatter}

%%
%% Start line numbering here if you want
%%
% \linenumbers

%% main text

\section{Introduction}\label{introduction}

The problem of testing whether a number is a quadratic residue modulo a composite $N$ is believed to be computationally hard. Various cryptographic protocols rely on this hardness assumption~\cite{Paper-BlumBlumShubPseudoRandom,Paper-CoinFlipping-Blum} and many applications, such as the Goldwasser-Micali cryptosystem~\cite{Paper-ProbEncryption-Goldwasser} require sampling uniformly distributed quadratic residues modulo an integer $N$. Moreover, this problem plays an important role in computational complexity, in particular, being the first such problem known to admit a zero-knowledge proof~\cite{Paper-KnowledgeComplexity-Goldwasser}.
%We define formally what a quadratic residue is and provide some number-theoretic background on Subsection~\ref{sec:numbertheory}.

An \emph{encoding} is a representation of objects from a finite set. Although there are various ways of representing these objects, in this paper we are interested in assigning, for each object, a positive integer in a given range. When this range size is equal to the size of the set being encoded, we say that the encoding is an \emph{indexing}, the positive integer assigned to the object is an \emph{index}, and the procedure that, given an index, outputs the object, is a \emph{decoding procedure}.

Some encodings are interesting even when the computation required for decoding is unfeasible or the corresponding set is infinite, such as the effective enumeration of Turing Machines in~\cite{Book-Kolmogorov-LiVitanyi}. By itself, the definition of an encoding does not deal with the time complexity required to decode an index. But, it is usually desirable for an encoding to be efficiently (i.e., polynomial time) decodable. There are, however, some caveats on how to come up with a precise definition for efficiency. Asymptotically, it makes sense to talk about \emph{ensembles of encodings}, each uniquely identified by a string $x$. So the polynomial time requirement should be in function of the size of $x$. This is the schema used in~\cite{Paper-MCSPMKTPGI-Allender}. In Section~\ref{sec:coding} we provide precise definitions for such concepts. 

In this paper we show an indexing for the set of quadratic residues modulo $N$ that is decodable in polynomial time on the size of $N$, when the factorization of $N$ is given. If such factorization is not given in the input, the procedure can be seen as a \emph{non-uniform polynomial time algorithm}, such as a circuit that has the factorization hardcoded. We note that in many applications regarding quadratic residues, usually $N = PQ$ for two primes $P$ and $Q$. Nevertheless the results presented here are applicable to any $N \in \mathbb{N}$. 

%TODO conferir
A consequence of the indexing presented here is the possibility of sampling uniformly distributed quadratic residues modulo $N$ in strict polynomial time and requiring the theoretical minimum amount of random bits. A previously known procedure for this same problem runs in expected (not strict) polynomial time and requires more random bits~\cite{Paper-KnowledgeComplexity-Goldwasser}. 

\section{Preliminaries}

%Quadratic Residues

\subsection{Number Theory}\label{sec:numbertheory}

Define the set $\mathbb{Z}_N^* = \{ 1 \leq x \leq N - 1 \mid \gcd(x,N) = 1\}$ for $N \in \mathbb{N}$. Given $N$ and its prime factorization $p_1^{k_1}\dots p_r^{k_r}$, the size of $\mathbb{Z}_N^*$ is given by Euler's totient function $\phi(N) = \prod_{i=1}^r (p_i -1)p_i^{k_i-1}$. $\mathbb{Z}_N^*$ forms a group under modular multiplication.

A \emph{quadratic residue modulo $N$} is an integer  $z$ such that $z \equiv x^2 \pmod{N}$ for an integer value of $x$. The \emph{set of quadratic residues modulo $N$} is defined as $\textrm{QR}(N) = \{z \in \mathbb{Z}_N^* \mid \exists x \in \mathbb{Z}_N^* \textrm{ s.t. }z \equiv x^2 \pmod{N}\}$. When $z \equiv x^2 \pmod{N}$ for $x \in \mathbb{Z}_N^*$, we call $x$ a \emph{square root} of $z$ modulo $N$.

When $N = n_1n_2\dots n_k$, where each of the $n_i$ are pairwise coprime, the Chinese Remainder Theorem establishes a group isomorphism between the groups $\mathbb{Z}_N^*$ and the direct product $\mathbb{Z}_{n_1}^* \times \mathbb{Z}_{n_2}^* \times \dots \times \mathbb{Z}_{n_k}^*$. This mapping is computable in time $o(\log^3{N})$, since it requires running the Extended Euclidean Algorithm, of time complexity $o(\log^2{N})$, at most $k \le \log{N}$ times (see for example~\cite[pp. 107-109]{Book-NumberTheory-Rosen}).

\subsection{Coding Theory}\label{sec:coding}

There are various notions of encoding in coding theory. Following~\cite{Paper-MCSPMKTPGI-Allender}, we define encodings from a set of integers to arbitrary sets via a decoder function $D$. Given $N \in \mathbb{N}$, let $[N]$ be the set $\{1,2,\dots,N\}$.

\begin{definition} (encoding and indexing). Let $S$ be a finite set. An \emph{encoding} of $S$ is a function $D : [I] \rightarrow S$ such that for every $s \in S$ there exists $i \in [I]$ such that $D(i) = s$. An \emph{indexing} is an encoding with $I = \size{S}$.
\end{definition}

An \emph{ensemble of encodings} $\{D_x\}$ is an infinite sequence of encodings, each uniquely identified by a string $x$. As discussed in Section~\ref{introduction}, we require the decoding algorithm to run in polynomial time on the size of $x$.

Note that there may be a polynomial time decoder algorithm that takes a polynomial size extra information, which may not be computable in polynomial time in the size of $x$. That is our case, since our decoding algorithm requires knowledge of the factorization of $N$. This notion is captured by non-uniform computation (algorithms that take advice or, equivalently, circuit families).

Following~\cite{Paper-MCSPMKTPGI-Allender}, we say that an ensemble of encodings $\{D_x\}$ is \emph{decodable by polynomial size circuits} if for each $x$ there is a circuit of size $\poly{\size{x}}$ that computes $D_x(i)$ for every $i \in [N_x]$. In the case where the function $(x,i) \mapsto D_x(i)$ is (uniformly) computable in time $\poly{\size{x}}$, we call the ensemble \emph{uniformly decodable in polynomial time}. Note that there may be a different circuit for each $x$ that indexes the ensemble, which differentiates this definition from the usual definition of circuit families, where there may be a different circuit for each input size.

Examples of indexings that are uniformly decodable in polynomial time are the Lehmer Code for permutations (see for example~\cite[pp. 12-13]{Book-Knuth-ACP3}) and the encodings of cosets of permutation subgroups presented by~\cite{Paper-MCSPMKTPGI-Allender}. An example of an encoding that, up to this date, is decodable by polynomial size circuits however is not known to be uniformly decodable in polynomial time is the one implicit in the Encoding Lemma of~\cite{Paper-MCSPMKTPGI-Allender-tr}.

\section{Indexing Quadratic Residues}

In this section we present our main contribution, an indexing for the set of quadratic residues modulo any $N \in \mathbb{N}$ that is decodable by polynomial size circuits. In our demonstration we use several classical results regarding quadratic residues, these can be found in~(\cite[pp. 63-71]{Book-DA-Gauss-tr}). Formally, we present an ensemble of encodings $\{D_N\}$, indexed by $\langle N \rangle$, the binary representation of a natural number $N$, such that $D_N : [I_N] \rightarrow \textrm{QR}(N)$, where $I_N = \size{\textrm{QR}(N)}$, that is decodable by polynomial size circuits.

We now present Proposition~\ref{prop-encoding} involving mixed radix encoding (for further details see \cite[pp. 327]{Book-Knuth-ACP2}). 

\begin{proposition}\label{prop-encoding}(\textbf{Mixed Radix Encoding}) Let $x_1,x_2,\dots,x_r$ and $p_1,p_2,\dots,p_r$ be natural numbers such that $x_i < p_i$ for $1 \leq i \leq r$, let also $N = p_1p_2\dots p_r$. There is a way to encode all of $x_1,x_2,\dots,x_r$ into a single natural $w \in [N]$ such that each $x_i$ can be recovered, given $w$ and the values of $p_1,p_2,\dots,p_r$, in time $o(\log^3{N})$.
\end{proposition}

%TODO conferir
Note that since there are exactly $p_1p_2\dots p_r = N$ possibilities for the values of $(x_1,x_2,\dots,x_r)$, Proposition~\ref{prop-encoding} actually establishes a bijection between $[N]$ and the possible values of $(x_1,x_2,\dots,x_r)$. We also present a restatement of Hensel's Lemma restricted to quadratic residues (see for example~\cite[pp. 179-183]{Book-CommutativeAlgebra-Eisenbud}).

\begin{lemma} (\textbf{Hensel's Lemma, restated}) Let $p$ be a prime number and $z \equiv x^2 \pmod p$. For all $k > 1$, there exists $y \in Z^{*}_{p^k}$, such that $z \equiv y^2 \pmod{p^k}$ and $x \equiv y \pmod p$.  
\end{lemma}

Next, in Theorem~\ref{theo-main}, we state our main result.

%There exists an indexing for every set of quadratic residues modulo a natural $N$ that is decodable by polynomial size circuits.

\begin{theorem}\label{theo-main} There is an ensemble of encodings $\{D_N\}$, indexed by $\langle N \rangle$, the binary representation of a natural number $N$, such that $D_N : [I_N] \rightarrow \textrm{QR}(N)$ for $I_N = \size{\textrm{QR}(N)}$, that is decodable by polynomial size circuits. 
\end{theorem}

First we present the general proof idea, then formalize it. We observe that to retrieve a quadratic residue $z \in \textrm{QR}(N)$ it suffices to know one square root of $z$ modulo each of the distinct prime powers dividing $N$. These square roots can then be recombined through the Chinese Remainder Theorem to obtain a square root $x \in \mathbb{Z}_N^*$ of $z$, which is then squared modulo $N$ to obtain $z$. 

If the factor is a power of $2$, such as $2^k$, let $y \in \mathbb{Z}_{2^k}^*$ be a square root of $z$ modulo $2^k$. In case $k \leq 3$, there is only one quadratic residue, the number $1$, and we can hard code (on the decoder algorithm) $y = 1$ as a square root. When $k > 3$, there is a square root $y$ of $z$ such that $y < 2^{k-2}$ (since all such numbers are incongruent modulo $2^k$ when squared), and then there is only the need to know the value $c < 2^{k - 3}$ such that $y = 1 + 2c$, since $y$ is always odd.

On the other hand, if the factor is a power of an odd prime number $p_i^{k_i}$, we need to know a square root $y_i \in \mathbb{Z}_{p_i^{k_i}}^*$ of $z$ modulo $p_i^{k_i}$. Modulo $p_i$, there will always be a square root $x_i$ of $z$ such that $x_i \leq (p_i - 1)/2$, in case $k_i = 1$, this information suffices. However, in case $k_i > 1$, we need more information. In this case, by Hensel's Lemma, there exists a square root of $y_i$ modulo $p_i^{k_i}$ of $z$ such that $x_i \equiv y_i \pmod{p_i}$. Then we have $y_i = x_i + c_ip_i$ for $c_i < p_i^{k_i - 1}$. It suffices then to know both $x_i$ and $c_i$ to recover $y_i$.

\begin{pot} We present an encoding $D_N : [I_N] \rightarrow \textrm{QR}(N)$ for $I_N = \size{\textrm{QR}(N)}$ and a polynomial size circuit that computes $D_N(Z)$ for an index $Z \in [I_N]$. Let $N = 2^kp_1^{k_1}\dots p_r^{k_r}$ be the prime factorization of $N$, where $k \geq 0$, each $p_i$ is a distinct odd prime and $k_i \geq 1$ for all $i$. Let also $z \in \textrm{QR}(N)$.

First we analyze the case where $N$ is an odd number (i.e. $k = 0$). Let  $x_i$ and $c_i$ for $1 \leq i \leq r$ be the numbers in the discussion above. Also from the same discussion, note that these values, together with the factorization of $N$, allow us to recover the quadratic residue $z$. We encode the values of $x_i - 1$, since $x_i \geq 1$, and $c_i$ for all $i$ into a single value $Z \in [I_N]$ using mixed radix encoding. Since  $x_i - 1 < (p_i - 1)/2$ and $c_i < p_i^{k_i - 1}$ for all $i$, we have
\begin{align*}
    Z   &\leq \prod_{i = 1}^{r} \left (\frac{p_i - 1}{2} \right )p_i^{k_i - 1} \\
        &= \frac{1}{2^r}\prod_{i = 1}^{r} (p_i - 1)p_i^{k_i - 1} \\
        &= \frac{\phi(N)}{2^r},
\end{align*}
%Troquei de circuit pra algorithm
which is precisely the size of $\textrm{QR}(N)$ for odd $N$. This also applies for the case where $k \leq 3$, since the value of $y$ is fixed to $1$ by the algorithm, and there is no need to store it in $Z$.

In case $N$ is an even number and $k > 3$, we also encode into the value of $Z$ the value of $c < 2^{k - 3}$, and then
\begin{align*}
    Z   &\leq 2^{k - 3}\prod_{i = 1}^{r} \left (\frac{p_i - 1}{2} \right )p_i^{k_i - 1} \\
        &= \frac{1}{2^{r + 2}}2^{k - 1}\prod_{i = 1}^{r} (p_i - 1)p_i^{k_i - 1} \\
        &= \frac{\phi(N)}{2^{r + 2}},
\end{align*}
which, again, is precisely the size of $\textrm{QR}(N)$ for even $N$ divisible by $2^k$.

The final step of the proof is to show that the ensemble $\{D_N\}$ is decodable by polynomial size circuits. We show that by providing a polynomial time decoder algorithm that receives as advice the complete factorization of $N$.

\begin{algorithm}[H]
\floatname{algorithm}{Decoder Algorithm}
\caption{- receives as advice the factorization of $N = 2^kp_1^{k_1}\dots p_r^{k_r}$ and as input an index $Z \in [I_N]$}
\label{algoritmo:decodificador}
\begin{algorithmic}[1]
    \STATE If $0 \leq k \leq 3$, let $y = 1$.
    \STATE Recover from $Z$ the values of $x_i$, $c_i$ (and $c$, when $k > 3$), using mixed radix encoding together with the values of $(p_i - 1)/2$ and $p_i^{k_i - 1}$ for all $i$.
    \STATE If $k > 3$, let $y = 1 + 2c$.
    \STATE Let $y_i = x_i + c_ip_i$ for all $i$.
    \STATE Recover $x \in \mathbb{Z}_N^*$ using the Chinese Remainder Theorem and the values of $y_i$ for all $i$ (and $y$ when $k > 3$).
    \STATE Output $x^2 \pmod{N}$
\end{algorithmic}
\end{algorithm}

Steps $2$ and $4$ require at most $\log{N}$ multiplications or divisions, which run in time $o(\log^2{N})$. Steps $2$ takes time $o(\log^3{N})$ by Proposition~\ref{prop-encoding}. Step $5$ also takes time $o(\log^3{N})$ to run the chinese remainder algorithm. The other steps are easily seen to be of lower time complexity. Therefore, the running time of the decoder algorithm is bounded by $o(\log^3{N})$. \qed
\end{pot}

%TODO conferir
\section{A Consequence of the Indexing}

The usual way to sample uniformly distributed quadratic residues is to randomly select a number $x$ between $1$ and $N - 1$, testing if $\gcd(x,N) = 1$ (repeating the process if the test fails), then squaring $x$ modulo $N$. This procedure is known to take expected polynomial time~\cite{Paper-KnowledgeComplexity-Goldwasser} and requires around $\log{N}$ random bits to obtain a sample, more than the information theoretical minimum of $\log{(\phi(N)/2^r)}$ for odd $N$ with $r$ distinct prime factors.

%sampling a distribution with entropy $\log{N}$ to obtain the desired, when $N$ is odd, $\log{(\phi(N)/2^r)}$ entropy distribution, where $r$ is the number of distinct prime factors of $N$.

Using the indexing presented in this paper, one can sample quadratic residues modulo $N$ when the factorization of $N$ is known by sampling a uniformly distributed number in $[\phi(N)/2^r]$ and running the decoder algorithm, this requires $\log{(\phi(N)/2^r)}$ random bits. This procedure attains the information theoretical minimum amount of randomness required to sample a uniformly distributed quadratic residue modulo $N$. Furthermore, this procedure allows for strict, instead of expected, polynomial time sampling of quadratic residues.

\section{Conclusion and Open Problems}

%TODO conferir
We have shown a non-uniform efficiently decodable indexing for the set of quadratic residues modulo any natural $N$, when the factorization of $N$ is known. While our objective is mainly in the information theoretical aspects of quadratic residues, there may be some practical consequences for sampling procedures. In many applications where quadratic residues sampling is necessary, the factors of $N$ are already known~\cite{Paper-ProbEncryption-Goldwasser,Paper-BlumBlumShubPseudoRandom}. In such cases, our procedure for generating random quadratic residues can be effectively applied.

It is a natural open question whether there exists an indexing that is uniformly and efficiently decodable. Since our construction relies on the knowledge of the factorization of $N$, and given the difficulty of the factoring and quadratic residuosity problems, it might be unlikely that such an indexing exists. Considering that, it would be interesting to directly relate the existence of such an indexing to the difficulty of these problems. Note, however, that even if there was such an indexing, it is not even clear whether it would allow for sampling of quadratic residues without the need to factor $N$, since until now there is no known efficient way to compute the size of $\textrm{QR}(N)$, that currently relies on knowing both $\phi(N)$ and the number of distinct prime factors of $N$.

\section{Acknowledgements}
The first author acknowledges a scholarship from the National Council for Scientific and Technological Development (CNPq).

%There are however, to the extent of our knowledge, no far reaching consequences, such as the possibility of testing whether a number is or not a quadratic residue in polynomial time in .

%In case it did, it is not even clear whether it would allow randomness efficient sampling of quadratic residues without the need to factor $N$, since until now there is no known efficient way to compute $\phi(N)$. 

%There are however, to the extent of our knowledge, no other far reaching consequences, such as the possibility of testing whether a number is or not a quadratic residue in polynomial time. Considering that, it would be interesting to directly relate the existence of such an indexing to the difficulty of the two mentioned problems.
%% References
%%
%% Following citation commands can be used in the body text:
%% Usage of \cite is as follows:
%%   \cite{key}         ==>>  [#]
%%   \cite[chap. 2]{key} ==>> [#, chap. 2]
%%

%% References with bibTeX database:
\bibliographystyle{elsarticle-num}
% \bibliographystyle{elsarticle-harv}
% \bibliographystyle{elsarticle-num-names}
% \bibliographystyle{model1a-num-names}
% \bibliographystyle{model1b-num-names}
% \bibliographystyle{model1c-num-names}
% \bibliographystyle{model1-num-names}
% \bibliographystyle{model2-names}
% \bibliographystyle{model3a-num-names}
% \bibliographystyle{model3-num-names}
% \bibliographystyle{model4-names}
% \bibliographystyle{model5-names}
% \bibliographystyle{model6-num-names}

%\bibliography{references.bib}

\begin{thebibliography}{10}
\expandafter\ifx\csname url\endcsname\relax
  \def\url#1{\texttt{#1}}\fi
\expandafter\ifx\csname urlprefix\endcsname\relax\def\urlprefix{URL }\fi
\expandafter\ifx\csname href\endcsname\relax
  \def\href#1#2{#2} \def\path#1{#1}\fi

\bibitem{Paper-BlumBlumShubPseudoRandom}
L.~Blum, M.~Blum, M.~Shub, Comparison of two pseudo-random number generators,
  in: Advances in Cryptology: Proceedings of CRYPTO '82, Plenum, 1982, pp.
  61--78.

\bibitem{Paper-CoinFlipping-Blum}
M.~Blum, Coin flipping by telephone a protocol for solving impossible problems,
  SIGACT News 15~(1) (1983) 23--27.

\bibitem{Paper-ProbEncryption-Goldwasser}
S.~Goldwasser, S.~Micali, Probabilistic encryption, Journal of Computer and
  System Sciences 28~(2) (1984) 270 -- 299.

\bibitem{Paper-KnowledgeComplexity-Goldwasser}
S.~Goldwasser, S.~Micali, C.~Rackoff, The knowledge complexity of interactive
  proof systems, SIAM J. Comput. 18~(1) (1989) 186--208.

\bibitem{Book-Kolmogorov-LiVitanyi}
M.~Li, P.~VitÃ¡nyi, An introduction to Kolmogorov complexity and its
  applications, 2nd Edition, Springer-Verlag, 1997.

\bibitem{Paper-MCSPMKTPGI-Allender}
E.~Allender, J.~A. Grochow, D.~van Melkebeek, C.~Moore, A.~Morgan, {Minimum
  Circuit Size, Graph Isomorphism, and Related Problems}, in: A.~R. Karlin
  (Ed.), 9th Innovations in Theoretical Computer Science Conference (ITCS
  2018), Vol.~94 of Leibniz International Proceedings in Informatics (LIPIcs),
  Schloss Dagstuhl--Leibniz-Zentrum fuer Informatik, Dagstuhl, Germany, 2018,
  pp. 20:1--20:20.

\bibitem{Paper-MCSPMKTPGI-Allender-tr}
E.~Allender, J.~A. Grochow, D.~van Melkebeek, C.~Moore, A.~Morgan, Minimum
  circuit size, graph isomorphism, and related problems, Tech. Rep. TR17-158,
  Electronic Colloquium on Computational Complexity {(ECCC)} (2017).

\bibitem{Book-NumberTheory-Rosen}
K.~Rosen, Elementary Number Theory and Its Applications, Pearson, 2011.

\bibitem{Book-Knuth-ACP3}
D.~E. Knuth, The Art of Computer Programming, Volume 3: Sorting and Searching,
  Addison-Wesley., 1973.

\bibitem{Book-DA-Gauss-tr}
W.~Waterhouse, J.~Brinkhuis, A.~Clarke, C.~Gauss, C.~Greiter, Disquisitiones
  Arithmeticae, Springer New York, 1986.

\bibitem{Book-Knuth-ACP2}
D.~E. Knuth, The Art of Computer Programming, Volume 2 (3rd Ed.): Seminumerical
  Algorithms, Addison-Wesley Longman Publishing Co., Inc., Boston, MA, USA,
  1997.

\bibitem{Book-CommutativeAlgebra-Eisenbud}
D.~Eisenbud, Commutative Algebra: With a View Toward Algebraic Geometry,
  Graduate Texts in Mathematics, Springer, 1995.

\bibitem{paper-PowerRandomStrings-Allender}
E.~Allender, H.~Buhrman, M.~KouckÃ½, D.~van Melkebeek, D.~Ronneburger, Power
  from random strings, SIAM Journal on Computing 35~(6) (2006) 1467--1493.

\bibitem{Paper-ZKandMCSP-Allender}
E.~Allender, B.~Das, Zero knowledge and circuit minimization, Information and
  Computation 256 (2017) 2 -- 8.

\bibitem{Paper-MKTPGIold-Allender}
E.~Allender, J.~A. Grochow, C.~Moore, Graph isomorphism and circuit size, Tech.
  Rep. TR15-162, Electronic Colloquium on Computational Complexity {(ECCC)}
  (2015).

\end{thebibliography}

\end{document}